\documentclass{aastex}
\usepackage{spr-astr-addons}
\usepackage{url}

\usepackage{graphicx}
\usepackage{esint}
\usepackage{epsfig}

\RequirePackage{color}

\makeatletter

\@ifundefined{textcolor}{}
{%
 \definecolor{BLACK}{gray}{0}
 \definecolor{WHITE}{gray}{1}
 \definecolor{RED}{rgb}{1,0,0}
 \definecolor{GREEN}{rgb}{0,1,0}
 \definecolor{BLUE}{rgb}{0,0,1}
 \definecolor{CYAN}{cmyk}{1,0,0,0}
 \definecolor{MAGENTA}{cmyk}{0,1,0,0}
 \definecolor{YELLOW}{cmyk}{0,0,1,0}
 }

\begin{document}

\title{Constraints on the Combined Models with $R^{2-q}$ Inflation and Viable $f(R)$ Dark Energy }
\shorttitle{$R^{2-q}$ Inflation and Viable $f(R)$ Dark Energy }
\shortauthors{Geng, Lee and Lin}

\author{Chao-Qiang~Geng$^{1,2,3}$,
Chung-Chi Lee$^{2}$,
and
Shan Lin$^{3}$
}
  \affil{$^{1}$Chongqing University of Posts \& Telecommunications, Chongqing, 400065, China\\
  $^{2}$Physics Division, National Center for Theoretical Sciences, Hsinchu,  Taiwan 300\\
  $^{3}$Department of Physics, National Tsing Hua University, Hsinchu,  Taiwan 300
  }


\begin{abstract}
We investigate the observational constraints on the modified gravity, which combines  the $R^{2-q}$ inflation with the power-law (exponential) type of 
the viable $f(R)$ dark energy models. We discuss the difference between the combined model and $R^{2-q}$ gravity in the inflationary epoch and obtain the constraints on the deviation power $q$ as well as  the parameters in $f(R)$ by using the CosmoMC package. 
The allowed ranges of the spectral index and tensor-to-scalar ratio from the Planck data are highly restricted, resulting 
in $q < 2.66 \times 10^{-2}$ and $2.17 \times 10^{-2}$ for the power-law and exponential types of $f(R)$ gravity, respectively.
\end{abstract}


\keywords{inflation; dark energy; modified gravity}


\section{Introduction}
Various cosmological observations such as Type-Ia Supernovae (SNIa)~\citep{Riess(1998), Perlmutter(1998)}, cosmic microwave 
background radiation (CMB)~\citep{Spergel(2003), Spergel(2006), Komatsu(2008), Komatsu(2010)} and 
baryon acoustic oscillation (BAO)\citep{Eisenstein(2005)} indicate that our universe is currently under an accelerating expansion, 
which is known as the dark energy problem. On the other hand, there are some cosmological problems in which the hot Big Bang model cannot be explained, such as the horizon, flatness and monopole problems.
A rapid accelerating expansion era, known as inflation, is introduced right after the beginning of the universe to provide a graceful solution to these 
problems~\citep{Lyth(1998)}.
As a result, our universe has gone through two different eras of the accelerating expansion: inflation and late-time acceleration.

It is well known that the simplest way to realize the late-time acceleration is to introduce a cosmological constant into the Einstein equation.
The corresponding theory is referred to as the $\Lambda$CDM model,
which
fits with the observational data very well, and becomes the standard model of cosmology.
However, there exists several unnatural properties in this model, such as coincident and hierarchy problems.
In order to remove these problems, people have tried various methods. 
One of which is to introduce a homogeneous and isotropic energy density with a negative pressure into the theory of General Relativity (GR), 
and the other one is to modify the Einstein's gravity theory~\citep{Copeland(2006)}.
Among them,  $f(R)$ gravity is  one of not only the minimal extension of the $\Lambda$CDM model,
 but also  the most efficient theory to explain  inflation and dark energy.

To describe the late-time accelerating phenomenon, several viable $f(R)$
viable models have been proposed to satisfy
the constraints from theoretical considerations and cosmological observations~\citep{DeFelice(2010),Bamba(2011)}.
These viable $f(R)$ models can be classified into power-law and exponential types~\citep{Lee(2012), Geng(2014)}.
Under these viable conditions, the future singularities have been removed
and the instability of cosmological perturbation never occurs in the universe. 
Additionally, there still exists a broad window of experimental tests by space gravitational wave detections~\citep{Yang:2011cp}.
Through the Markov chain Monte Carlo method (MCMC) analysis, the viable $f(R)$ dark energy models get a smaller likelihood 
than that in the $\Lambda$CDM model, denoting that they are preferred by cosmological observations~\citep{Motohashi:2012wc, Geng(2014)}.
Besides the dark energy problem,  $f(R)$ gravity is also used to address the dark matter problem~\citep{Cembranos(2009), Cembranos(2011), 
Sharif(2014a), Sharif(2014b), Sharif(2015a),Sharif(2015b)} as well as the inflation scenario. 
In this manuscript, we will focus on the viable $f(R)$  models with dark energy  and inflation.

It is well-known that the $R^2$ inflation, first proposed by Starobinsky~\citep{Starobinsky(1980)}, 
successfully describes the reheating period~\citep{Vilenkin(1985), Mijic(1986), Ford(1986)} and predicts a tiny tensor-to-scalar ratio, 
which is consistent with the observations today~\citep{Ade(2015)}.
Since both $f(R)$ gravity and $R^2$ inflation achieve their goals by modifying the action of general relativity to a function of Ricci scalar, it is tempting to consider their actions together to provide a unified framework for both inflation and the late-time acceleration epoch.
In fact, the $R^2$ model remains one of the most attractive models today describing inflation because it solves the singularity problems when combined with $f(R)$ dark energy models~\citep{Appleby(2010), Lee(2012)}.

As a simple generalization of the $R^2$ inflation, 
many people~\citep{Martin(2014a), Martin(2014b), Martin(2014c), Codello(2014),  Rinaldi(2014c), Rinaldi(2014b), Rinaldi(2014c), 
Motohashi(2014), Myrzakul(2015), 
Nojiri(2003), Nojiri(2006),Nojiri(2007),Nojiri(2008),Cognola(2008),Elizalde(2011),Bamba(2008),Bamba(2013),Bamba(2014),Bamba(2015),
Chakravarty(2013), Chakravarty(2014)} have investigated the possibility of  the power in $R^2$ to deviate slightly from two, named the $R^{2-q}$ inflation, where $q \ll 1$. 
The $R^{2-q}$ inflation has been shown to be compatible with observations and predicts a larger scalar-to-tensor ratio than that in the $R^2$ inflation. 
It is thus natural to combine the $R^{2-q}$ inflation model with viable $f(R)$ dark energy models to ensure the late-time acceleration  of our universe.
As an example, a recent attempt of combining these two has been given in Ref.~\citep{Artymowski(2014)}.
In this paper, we would like to study the constraints on the combined inflation with dark energy models.
The action includes the early universe $R^{2-q}$ inflation and late-time $f(R)$ dark energy of the Starobinsky~\citep{Starobinsky(2007)} and exponential gravity~\citep{Zhang(2005), Cognola(2008), Linder(2009), Bamba(2010)} models, which belong to the power-law and exponential types, respectively.

We modify the Code for Anisotropies in the Microwave Background (CAMB)~\citep{Lewis(1999)}
and the Cosmological MonteCarlo (CosmoMC) package~\citep{Lewis(2002)} to study the constraints on inflation as well as $f(R)$ dark energy models from
the cosmological observations, including those of
 the CMB data from Planck~\citep{Ade(2013)}
and WMAP~\citep{Hinshaw(2012)}, BAO data from
Baryon Oscillation Spectroscopic Survey (BOSS)~\citep{Anderson(2012)} and
SNIa data from Supernova Legacy Survey (SNLS)~\citep{Astier(2005)}.
Since the MGCAMB program (Modification of Growth with CAMB)~\citep{Hojjati(2011)} uses the parametrized framework to include $f(R)$ gravity into CAMB, we only consider the linear perturbation and assume that the background evolution is the same as the $\Lambda$CDM model.

This paper is organized as follows. In Sec.~\ref{sec:theory}, we first give a brief review on the $R^{2-q}$ inflation with slow-roll parameters, and then introduce the $f(R)$ modifications in the scalar and tensor perturbations. 
We also estimate the spectral index $n_s$ and tensor-to-scalar ratio $r$.
In Sec.~\ref{sec:constraints}, we show our results of the observational constraints on the combined models by using the CosmoMC package.
Finally, we present our conclusions in Sec.~\ref{sec:conclusion}.

\section{The Action and Parameters } \label{sec:theory}
Let us begin by considering the 4-dimentional general $f(R)$ action
\begin{eqnarray}
\label{eq:action_tot}
S = \frac{M^2_{\text{Pl}}}{2} \int d^4 x \sqrt{-g} f(R) + S_M \,,
\end{eqnarray}
where $M_{\text{Pl}} \equiv (8\pi G)^{-1/2}$ is the reduced planck mass, $f(R)$ is a general function of the Ricci scalar $R$, and $S_M$ is the matter action, including relativistic and non-relativistic components.
In this study, we consider $f(R)$ to be the form,
\begin{eqnarray}
\label{eq:fR_tot}
f(R) = R + F(R) + \frac{R^{2-q}}{M^{2-2q}}  \,,
\end{eqnarray}
where $R^{2-q}$ and $F(R)$ are responsible for the inflation and late time dark energy, respectively.
It is worth to mention that the $R^2$ model was claimed to not only  realize  inflation but also 
solve the dark matter problem~\citep{Cembranos(2009), Cembranos(2011)}. The later 
has also been discussed with the extended dark matter models  in the literature~\citep{Sharif(2014a), Sharif(2014b), Sharif(2015a),Sharif(2015b)}.

In general, the F(R) models, which satisfy the viable conditions~\citep{DeFelice(2010)} can be categorized into two classes, the power law and exponential types.
To investigate the generic features in these classes, we concentrate on
the exponential type  with the exponential form of  gravity~\citep{Zhang(2005), Cognola(2008), Linder(2009), Bamba(2010)} and the power-law
one, called the Starobinsky  model~\citep{Starobinsky(2007)}.

%
\begin{eqnarray}
\label{eq:lagr_star}
\mathrm{Starobinsky:}~ F(R) = -\lambda R_c \left[ 1- \bigg( 1+ \frac{R^2}{R^2_c} \bigg) ^{-n} \right] \,, \\
\label{eq:lagr_exp}
\mathrm{Exponential:}~  F(R) = -\beta R_c (1- e^{R/R_c}) \,,~~~~~~~~~~~~
\end{eqnarray}
where $\lambda$ and $\beta$ are the model parameters and $R_c$ represents the constant characteristic curvature.

\subsection{$R^{2-q}$ inflation}~\label{subsec:rp_inf}
\begin{center}
\begin{figure}[h]
\includegraphics[width=0.49 \linewidth]{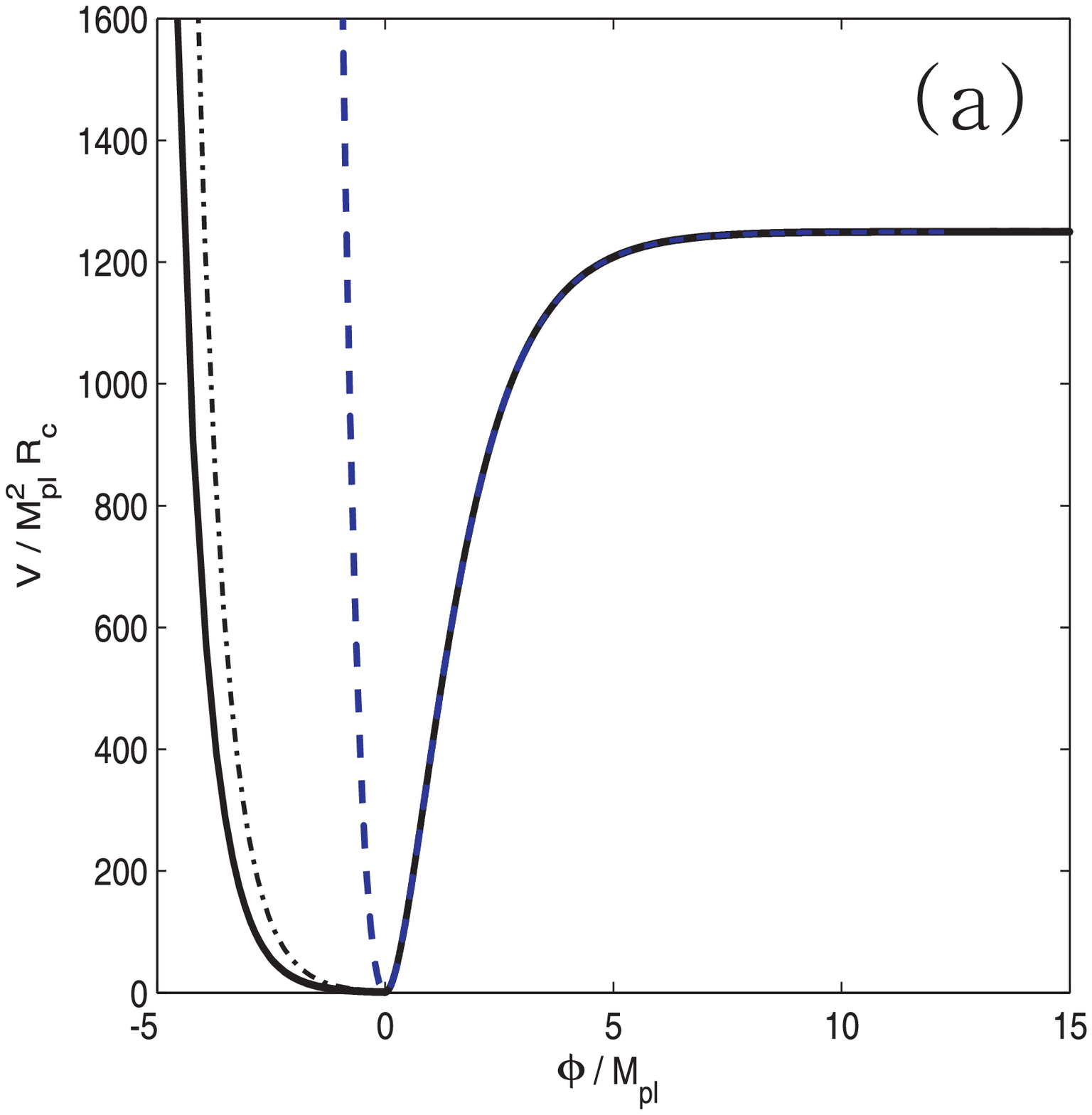}
\includegraphics[width=0.49 \linewidth]{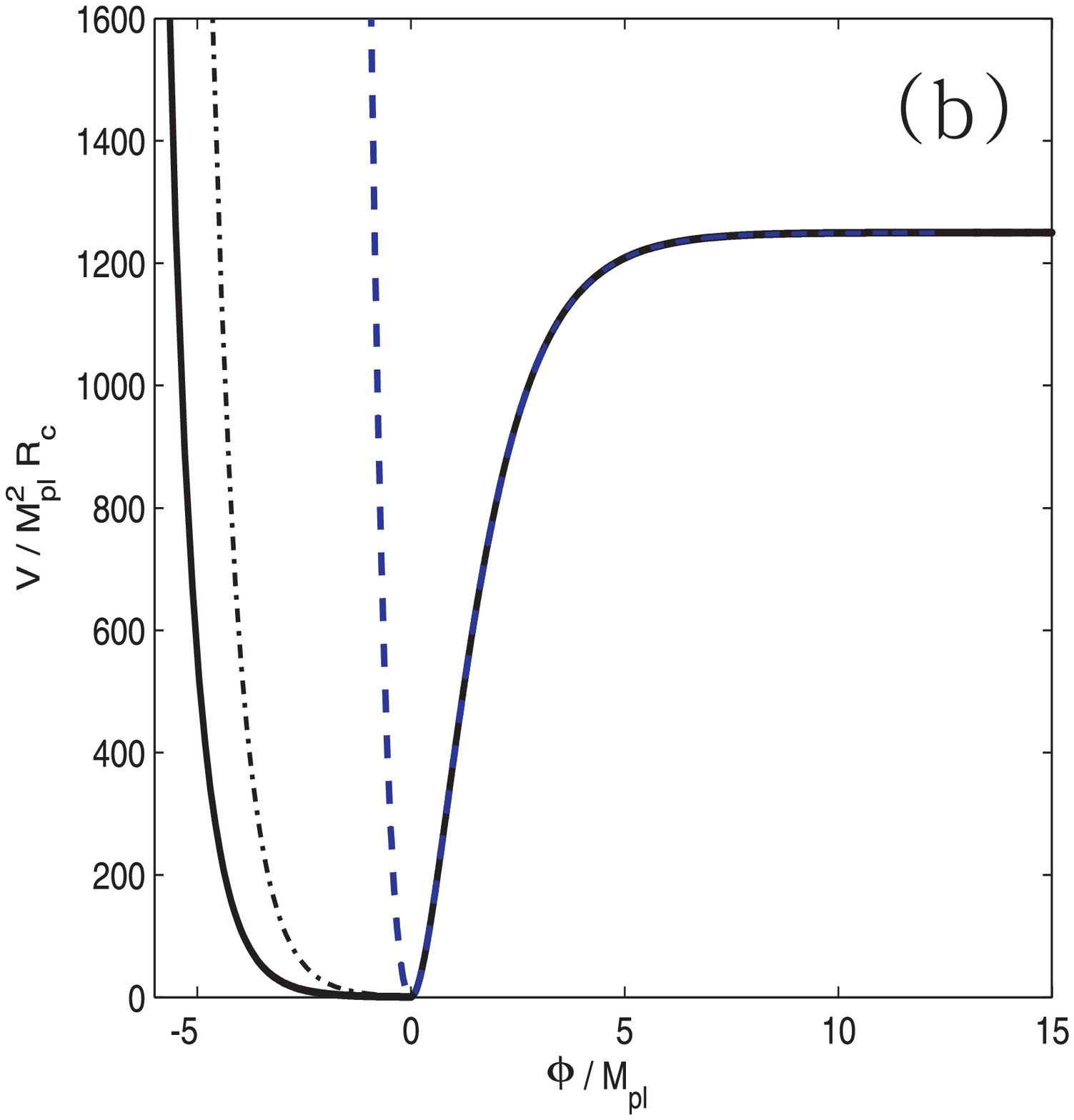}
\caption{
The potentials in the Einstein frame as functions of the scalar field $\phi$ 
in the combined model, $f(R)= R + F(R)+R^2/M^2$, 
where  $F(R)$ has the form of (a) the Starobinsky model with $(\lambda,\, n)=(2,\, 2)$  (solid line) and $(4,\, 2)$ (dash-dotted line) and (b) the exponential gravity with $\beta=2$ (solid line) and $\beta=4$ (dash-dotted line), while the dashed line corresponds to the $R^2$ inflation, $f(R) = R+ R^2/M^2$.}
\label{fg:5}
\end{figure}
\end{center}

\begin{center}
\begin{figure}[h]
\includegraphics[width=0.49 \linewidth]{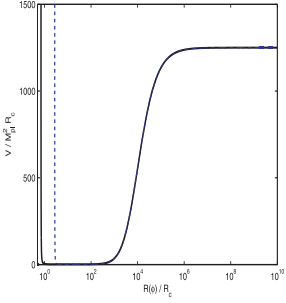}
\caption{
The relation of the potential in the Einstein frame and the curvature in the Jordan frame, 
where the solid and dashed lines correspond 
to the combined models with exponential ($\beta=2$) and  Starobinsky ($\lambda=2$ and $n = 2$) gravity, respectively.}
\label{fg:6}
\end{figure}
\end{center}

First of all, let us focus on the inflationary stage.
By conducting a conformal transformation $\tilde{g}_{\mu\nu} = f_R g_{\mu\nu}$ and $f_R \equiv \partial f(R) / \partial R = e^{\sqrt{2/3} ~ \phi / M_{\text{Pl}}}$, we transform the $f(R)$ action from the Jordan frame into Einstein frame,
\begin{eqnarray}
S = \int d^4x \sqrt{-\tilde{g}} \left[ \frac{M^2_{\text{Pl}}}{2}\tilde{R} - \frac{1}{2}\tilde{g}^{\mu\nu}\partial_{\mu}\phi\partial_{\nu}\phi - V(\phi) \right]\,.
\end{eqnarray}
where $\tilde{R}$ is the curvature in  the Einstein frame and the potential is given as
\begin{eqnarray}
\label{eq:potential}
V(\phi) = \frac{M^2_{\text{Pl}}}{2} \frac{R f_R - f}{f_R^2} \,.
\end{eqnarray}
Note that in the combined model, the potential depends on not only the $R^{2-q}$ term but also the $F(R)$ dark energy model, 
and the scalar field in the Einstein frame is defined by
\begin{eqnarray}
\label{eq:Ein_scalar}
f_R=e^{\sqrt{2/3}\phi/M_{\mathrm{pl}}}= 1+(2-q)\frac{R^{1-q}}{M^{2-2q}} +\frac{d F}{d R} \,.
\end{eqnarray}
From Eqs.~(\ref{eq:lagr_star}), (\ref{eq:lagr_exp}) and (\ref{eq:Ein_scalar}), we see that 
the large curvature regime corresponds to a large $\phi$, so that the behavior of 
the scalar field is dominated by the $R^{2-q}$ term, whereas the small curvature regime represents a small or negative $\phi$, resulting in the $F(R)$ term responsible for the evolution of the scalar.
Due to the non-linear form of Eqs.~(\ref{eq:lagr_star}) and (\ref{eq:lagr_exp}), it is not easy to derive the explicit form of $V(\phi)$ in the combined model. 
In Fig.~\ref{fg:5}, we numerically reverse Eq.~(\ref{eq:Ein_scalar}) to obtain the relation $R=R(\phi)$ and
show the potentials as functions of the scalar field by substituting $R(\phi)$ into Eq.~(\ref{eq:potential})
with the $R^2$ inflation and  combined model.
Note that, as an illustration, we simplify the model with $(q,\,M^2/R_c)=(0,\,10^4)$ and 
$F(R)$ is the Starobinsky  model in Eq.~(\ref{eq:lagr_star}) (left panel) with $(\lambda,\, n)=(2,\,2)$ (solid line) and $(4,\,2)$ (dash-dotted line), and the exponential gravity in Eq.~(\ref{eq:lagr_exp}) (right panel) with $\beta=2$ (solid line) and $4$ (dash-dotted line).
Obviously, when the curvature is large enough and close to the inflationary stage, the potentials of the combined models (solid line) mimic that of the $R^2$ inflation (dashed line).
In the small curvature regime ($\phi \lesssim 0^+$), the potentials of the $R^2$ inflation and the combined models increase to infinity simultaneously.
In the slow-roll regime, the $R^{2-q}$ inflation model is no difference from the combined model.
 However, the post inflationary history in the combined models  could have a significant change of that in the $R^2$ inflation.
On the other hand, in the post inflationary epoch, the curvature oscillation in the Jordan frame also exhibits a different behavior between the $R^2$ inflation and the combined models.
In the $R^2$ inflation, the scalaron oscillates around the potential minimum at $\phi=0$ in the Einstein frame, corresponding to the curvature $R$ oscillating around $R=0$ in the Jordan frame. 
On the contrary, the curvature oscillation in the combined models ``never'' reaches zero and evolves to be negative. As shown in Fig.~\ref{fg:6}, one has
\begin{eqnarray}
V\left( \phi(R) \right) \rvert_{R \rightarrow R_{cr}} \rightarrow \infty \,,
\end{eqnarray}
where $(\lambda,\, n)=(2,\,2)$ for the Starobinsky  model (dashed line) and $\beta=2$ for the exponential one (solid line).
We choose the exponential $F(R)$ gravity to be the example.
The potential in the Einstein frame is given from Eq.~(\ref{eq:potential}).
When the curvature is small enough and  $R/M^2$ becomes negligible, $V(\phi)$ approaches infinity at $f_R \simeq 1 + F_R \rightarrow 0$, i.e., 
\begin{eqnarray}
R \rightarrow R_{cr}= R_c \ln \beta \,,
\end{eqnarray}
which represents a positive critical curvature.

As discussed above, the post inflationary epoch history might be changed by including $F(R)$ dark energy component. 
However, 
matter and dark energy components, corresponding to $S_M$ and $F(R)$ in Eq.~(\ref{eq:fR_tot})  still can be ignored during the inflationary stage, respectively, so that the inflation potential is given by substituting $f(R) = R +\alpha R^{2-q}$ into Eq.~(\ref{eq:potential}),
\begin{eqnarray}
V(\phi) = V_0 e^{-2\sqrt{\frac{2}{3}} \frac{\phi}{M_{\text{Pl}}}} 
\left( e^{\sqrt{\frac{2}{3}} \frac{\phi}{M_{\text{Pl}}}} -1 \right) ^{\frac{2-q}{1-q}} \,,
\end{eqnarray}
where $V_0 \equiv (1-q) (2-q)^{(2-q)/(q-1)} \alpha^{1/(q-1)} M^2_{\text{Pl}}/2$ is a mass dimension-four constant.

\begin{center}
\begin{figure}[h]
\includegraphics[width=0.49 \linewidth]{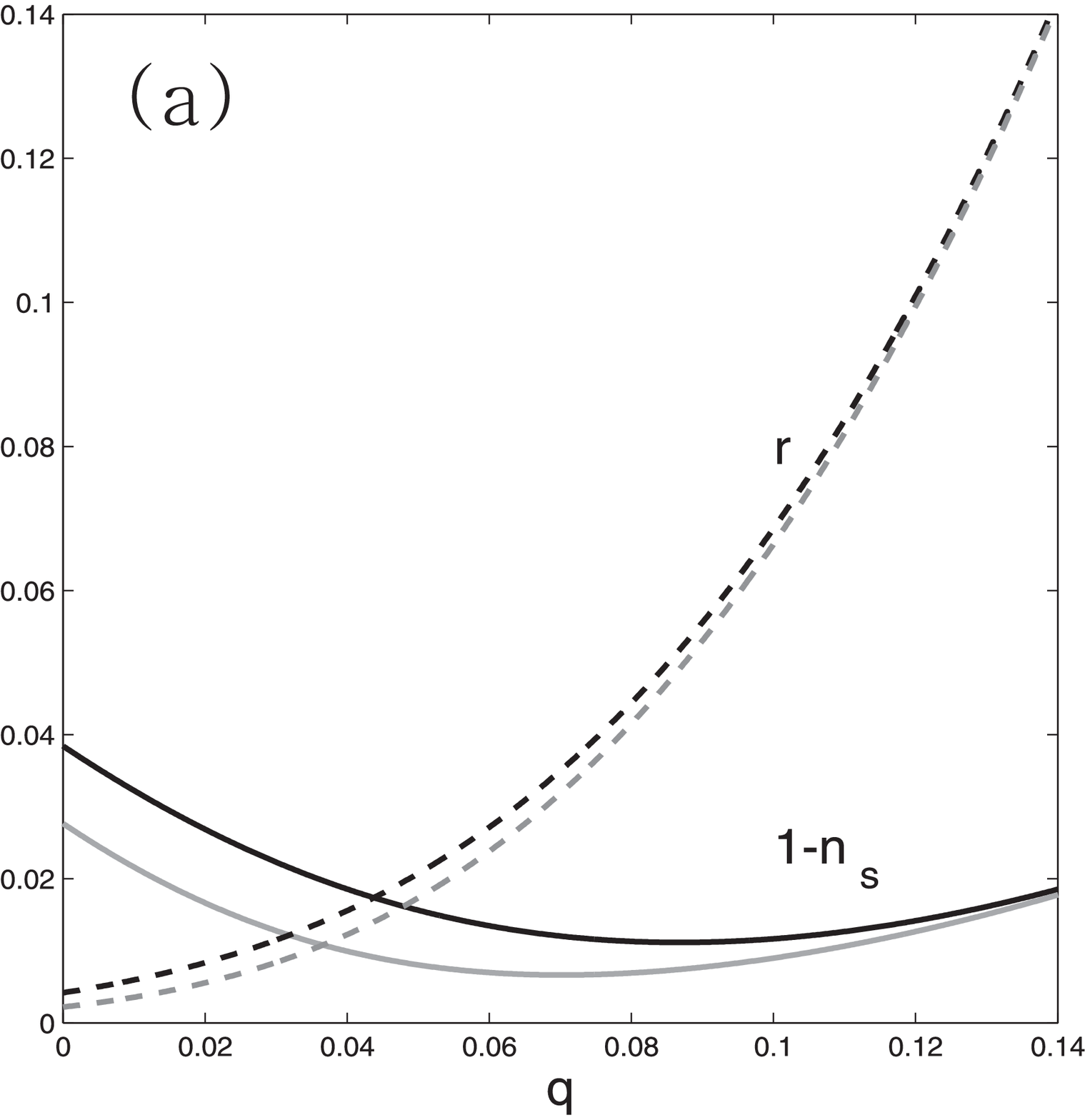}
\includegraphics[width=0.49 \linewidth]{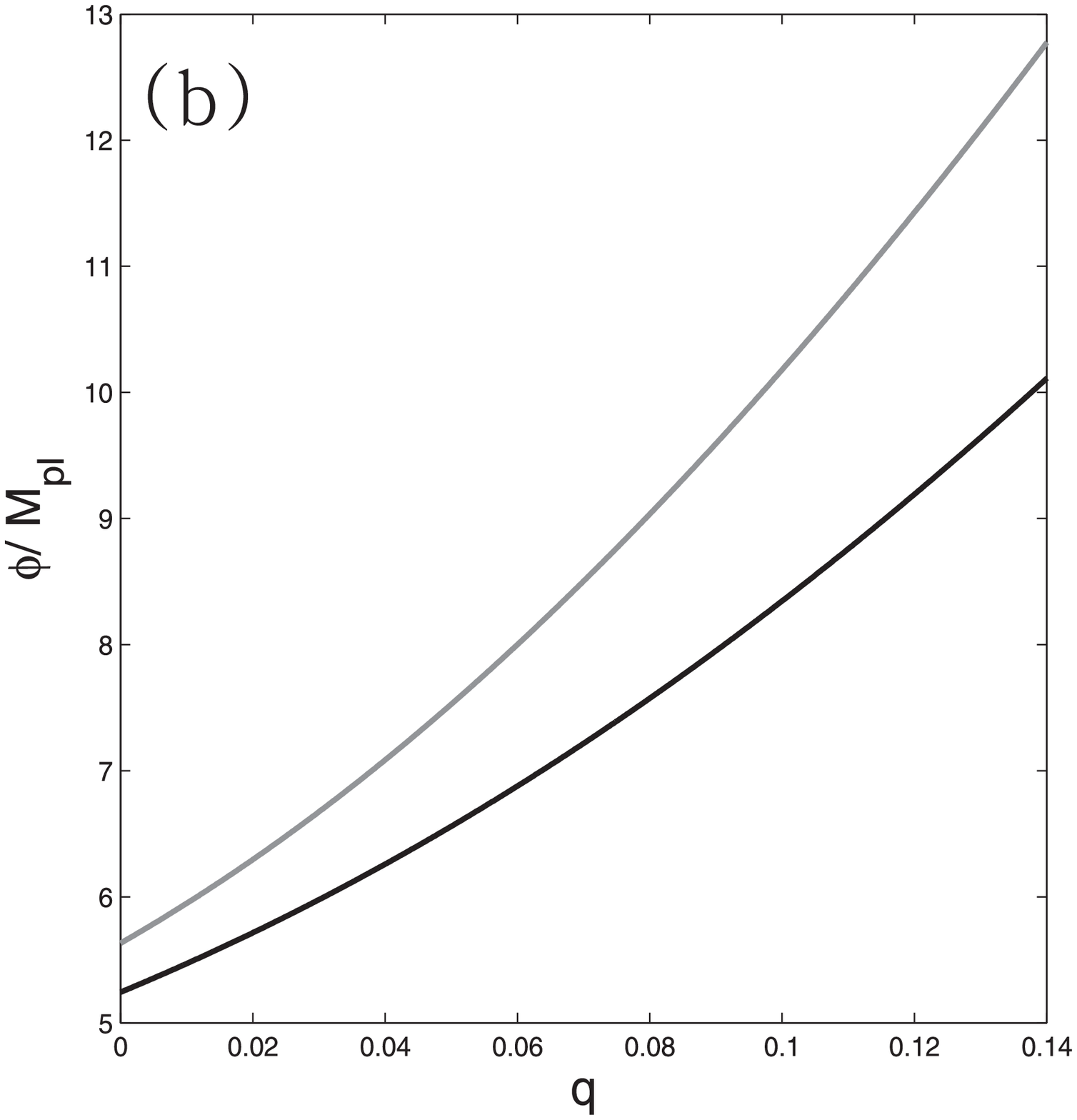}
\\
\includegraphics[width=0.49 \linewidth]{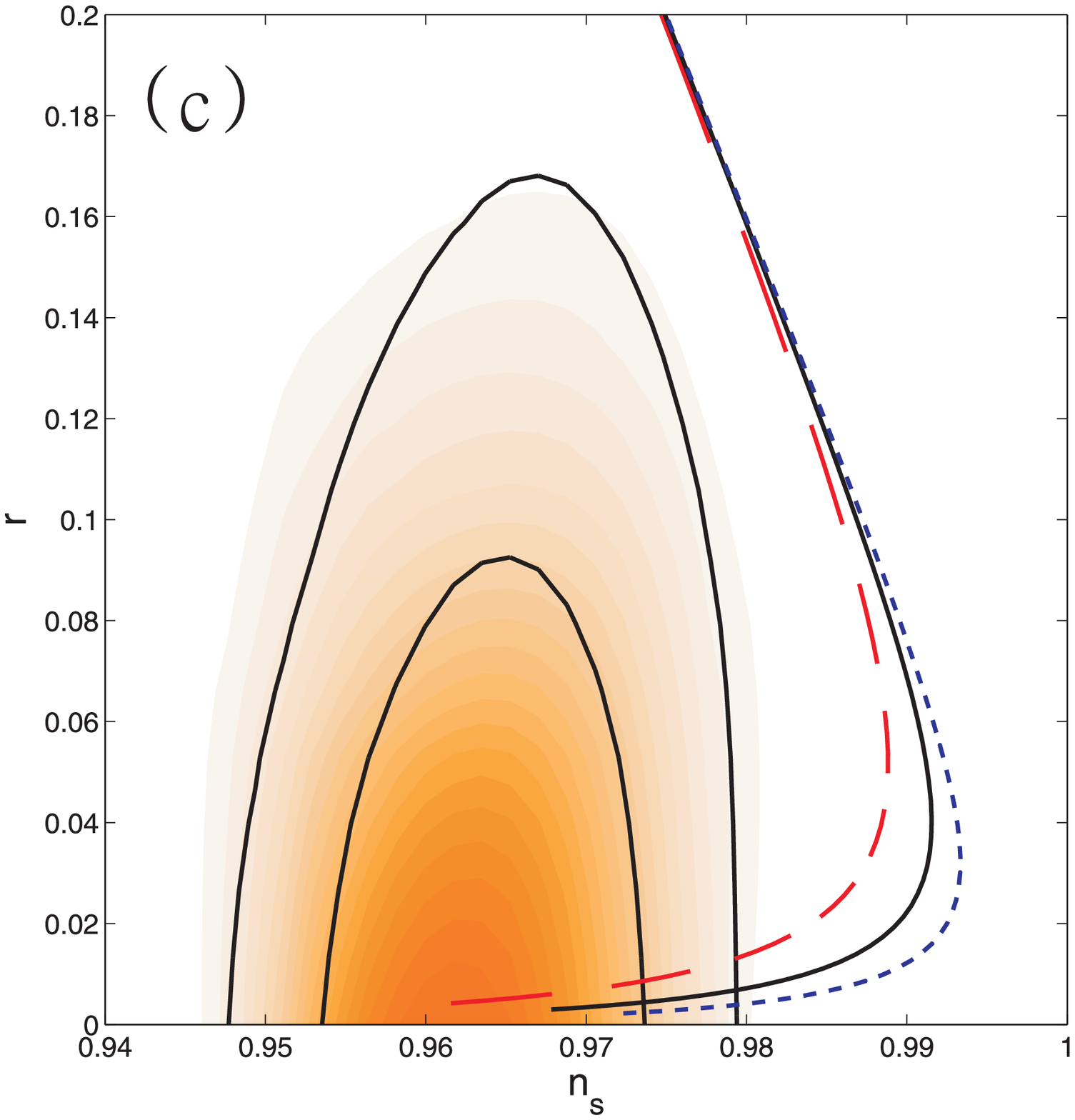}
\caption{
(a) The scalar spectral index $n_s$ (solid line) and tensor-to-scalar ratio $r$ (dashed line) as functions of the inflation power parameter $q$ in the $R^{2-q}$ inflation model, where the black and gray lines correspond to $N=50$ and $70$, respectively.
(b) The scalar at initial time, $\phi$, as functions of $q$ with $N=50$ (black line) and $70$ (gray line).
(c) The tensor-to-scalar ratio $r$ as a function of the scalar spectral index $n_s$, where the long-dashed, solid and dashed lines denote $N=50$, $60$ and $70$, while the contour plot presents the $1\sigma$ and $2\sigma$ bound in the $\Lambda$CDM model, respectively.}
\label{fg:1}
\end{figure}
\end{center}

Given the potential $V(\phi)$, we obtain the slow-roll parameters, defined by
\begin{eqnarray}
\label{eq:slow-roll-parm}
\epsilon \equiv \frac{M^2_{\text{Pl}}}{2} \left( \frac{V_{\phi}}{V} \right)^2, ~ \eta \equiv M^2_{\text{Pl}} \frac{V_{\phi \phi}}{V},~ 
 \xi \equiv M^ 4_{Pl} \frac{V_{\phi}V_{\phi \phi \phi}}{V^2}\,.
\end{eqnarray}
With these parameters, we can write down the inflation observables, such as the scalar and tensor spectral index ($n_s, n_t$), tensor-to-scalar ratio ($r$) and scalar spectral index running ($\alpha_s \equiv dn_s/d\ln k$) to be
\begin{eqnarray}
\label{eq:inf_parm1}
&&n_s - 1 = -6\epsilon + 2\eta \,, \\
\label{eq:inf_parm2}
&&n_t = - 2\epsilon \,, \\
\label{eq:inf_parm3}
&&r = 16\epsilon \,, \\
\label{eq:inf_parm4}
&&\alpha_s = 16\epsilon \eta -24\epsilon^2 -2\xi \,,
\end{eqnarray}
respectively.
The end of  inflation is usually assumed to satisfy the condition,
\begin{eqnarray}
\epsilon \lvert_{\phi=\phi_{end}} = 1 \,.
\end{eqnarray}
Therefore, one obtains
\begin{eqnarray}
\label{eq:phif}
\frac{\phi_{end}}{M_{\text{Pl}}} = \sqrt{\frac{3}{2}} \ln \left[ \frac{(2+\sqrt{3})(1-q)}{\sqrt{3}-(1+\sqrt{3})q} \right]\,.
\end{eqnarray}
Now, we can derive the number of e-foldings during inflation,
\begin{eqnarray}
\label{eq:efolding}
N \equiv \int^{\tilde{t}_{end}}_{\tilde{t}} \tilde{H} d\tilde{t} \simeq \frac{1}{M^2_{\text{Pl}}} \int_{\phi_{end}}^{\phi} \frac{V}{V_{\phi}} d\phi \,,
\end{eqnarray}
where the tilde denotes the variable in the Jordan frame. From the conformal transformation, we have $\tilde{H} d \tilde{t} = Hdt\left[1+\dot{f}_R/(2Hf_R)\right] \simeq Hdt$ during the slow-roll regime. As a result, the numbers of the e-folding in both Einstein and Jordan frames are the same, i.e. $N_E \simeq N$.
Substituting Eq.~(\ref{eq:phif}) into Eq.~(\ref{eq:efolding}), we find
\begin{eqnarray}
\label{eq:Nq1}
q \neq 0: ~~~~~~~~~~~~~~~~~~~~~~~~~~~~~~~~~~~~~~~~~~~~~~~~~~~
\nonumber\\
  N(\phi) = \frac{3(2-q)}{4q} \ln \left[ \frac{q e^{\sqrt{\frac{2}{3}}\frac{\phi}{M_{\text{Pl}}}}+2(1-q)}{q e^{{\sqrt{\frac{2}{3}}\frac{\phi_{end}}{M_{\text{Pl}}}}}+2(1-q)} \right]
\nonumber\\
 - \frac{3}{4} \left(\sqrt{\frac{2}{3}}\frac{\phi}{M_{\text{Pl}}}-\sqrt{\frac{2}{3}}\frac{\phi_{end}}{M_{\text{Pl}}}\right) \,, \\
\label{eq:Nq2}
q = 0:  ~~~~~~~~~~~~~~~~~~~~~~~~~~~~~~~~~~~~~~~~~~~~~~~~~~~
\nonumber\\
 N(\phi) = \frac{3}{4} \left( e^{\sqrt{\frac{2}{3}}\frac{\phi}{M_{\text{Pl}}}}-e^{\sqrt{\frac{2}{3}}\frac{\phi}{M_{\text{Pl}}}} \right)
 ~~~~~~~~~~~~~~
\nonumber\\
 - \frac{3}{4}\left(\sqrt{\frac{2}{3}}\frac{\phi}{M_{\text{Pl}}}-\sqrt{\frac{2}{3}}\frac{\phi_{end}}{M_{\text{Pl}}} \right) \,.
\end{eqnarray}
In order to avoid the negative value inside the logarithm in Eq.~(\ref{eq:Nq1}), we concentrate on $q>0$ in this study.
We can reverse the equations to get
\begin{eqnarray}
\label{eq:phii}
\phi= \phi(q,N)\,.
\end{eqnarray}
The slow-roll parameters are given by
\begin{eqnarray}
\label{eq:slow-roll-parm1}
 \epsilon &=& \frac{[qf_R + 2(1-q)]^2}{3(1-q)^2(f_R-1)^2}, \\
\label{eq:slow-roll-parm2}
 \eta &=& \frac{2[q^2f_R^2-(1-q)(2-5q)f_R+4(1-q)^2]}{3(1-q)^2(f_R-1)^2}, \\
\label{eq:slow-roll-parm3}
 \xi &=& 4[qf_R+2(1-q)][q^3f_R^3 
\nonumber \\
 &&+(1-q)(1-2q)(2-5q)f_R^2
\nonumber \\
&&-(1-q)^2(10-17q)f_R
\nonumber \\
&&+8(1-q)^3][9(1-q)^4(f_R-1)^4]^{-1} \,,
\end{eqnarray}
where $f_R = e^{\sqrt{2/3} ~ \phi /M_{\text{Pl}}}$. 
We now numerically analyze the $R^{2-q}$ inflation by using Eqs.~(\ref{eq:Nq1})-(\ref{eq:slow-roll-parm3}).
In Fig.~\ref{fg:1}a, we depict the scalar spectral index $n_s$ (solid line) and tensor-to-scalar ratio $r$ (dashed line) 
as functions of the power deviation $q$
in the $R^{2-q}$ inflation model, where
the black and gray lines represent the inflation e-foldings with $N=50$ (black) and $70$ (gray), respectively.
In Fig.~\ref{fg:1}b, we plot $\phi$ as functions of $q$ with $N=50$ and $70$ to demonstrate the values 
of the scalar field at the initial time as functions of $q$ by Eq.~(\ref{eq:phii}).
In Fig.~\ref{fg:1}c, we show  $r$ as a function of  $n_s$, where 
where the long-dashed, solid and dashed lines denote $N=50$, $60$ and $70$, 
while the contour plot presents the $\Lambda$CDM model constrained at 68\% and 95\% confidence levels with the same datasets used in 
Sec.~\ref{sec:constraints}, respectively.
 Comparing these two plots, we observe that the $R^{2-q}$ inflation allows a larger value of $r$,
 but 
 $q \lesssim O(10^{-2})$.

\subsection{$F(R)$ Dark Energy} \label{subsec:fR_DE}
After inflation, the scalaron settles down in the global minimum of the potential.
Hence, the $R^{2-q}$ term in the general action is lost its importance and can be ignored in the evolution afterwards.
We then have the same action as the $F(R)$ action, which is
\begin{eqnarray}
S = \frac{M^2_{\text{Pl}}}{2} \int d^4 x \sqrt{-g} (R + F(R)) + S_m \,.
\end{eqnarray}
One can obtain the field equation by varying the action with respect to the metric, given by
\begin{eqnarray}
\label{eq:fieldeq}
&&F_R R_{\mu\nu} - \frac{1}{2}F(R)g_{\mu\nu} + (g_{\mu\nu}\Box - \nabla_{\mu} \nabla_{\nu})F_R 
\nonumber\\
&&= \frac{1}{M^2_{\text{Pl}}}T_{\mu\nu} \,,
\end{eqnarray}
where $F_R \equiv dF(R)/dR$, $\Box = g^{\mu\nu}\nabla_{\mu} \nabla_{\nu}$ is the d'Alembertian operator and $T_{\mu\nu}$ is the energy-momentum tensor defined by
\begin{eqnarray}
T_{\mu\nu} \equiv \frac{-2}{\sqrt{-g}} \frac{\delta S_m}{\delta g^{\mu\nu}} \,.
\end{eqnarray}

To deal with the linear perturbation in $f(R)$ gravity, let us first write down the Friedmann-Robertson-Walker (FRW) metric in the Newtonian gauge,
\begin{eqnarray}
\label{eq:FRW-scalar}
ds^2 = a^2(\tau) \left[-(1+2\Psi)d\tau^2 + (1-2\Phi)\delta_{ij}dx^i dx^j \right] \,,
\end{eqnarray}
where $a(\tau)$ is the scale factor, $\tau$ is the conformal time, and $\Psi$ and $\Phi$ are the scalar perturbations.
The perturbed energy-momentum tensor is
\begin{eqnarray}
&& T^0_0 = -(\rho_M + \delta \rho_M) \,, \\
&& T^0_i = -(\rho_M + P_M)v_{M,i} \,,
\end{eqnarray}
where $v_M$ is the velocity field. By following the similar procedure in Ref.~\citep{Hojjati(2011)}, we can derive the relationship between the metric potential to the perturbation of the matter density in the sub-horizon limit,
\begin{eqnarray}
\label{eq:Poisson-1}
&& \frac{k^2}{a^2}\Psi = -4\pi G_{eff}(k,a)\rho_M \delta_M \,, \\
\label{eq:Poisson-2}
&& \frac{\Phi}{\Psi} = \gamma(k,a) \,,
\end{eqnarray}
where $\delta_M \equiv \delta\rho_M/\rho_M+3Hav_M$ is the gauge-invariant matter density perturbation and
\begin{eqnarray}
&& G_{eff}(k,a) = \frac{G}{F_R} \frac{1+4\frac{k^2}{a^2}\frac{F_{RR}}{F_R}}{1+3\frac{k^2}{a^2}\frac{F_{RR}}{F_R}} \,, \\
&& \gamma (k,a) = \frac{1+2\frac{k^2}{a^2}\frac{F_{RR}}{F_R}}{1+4\frac{k^2}{a^2}\frac{F_{RR}}{F_R}} \,.
\end{eqnarray}
We incorporate the linear perturbation effects by introducing the effective gravitational constant $G_{eff}$ and the ratio of Newtonian gauge scalar potentials as defined above.

For completeness, we consider not only the scalar Newtonian potentials but also the tensor perturbation~\citep{Zhou(2014)}. The perturbation of the FRW metric is given by,
\begin{eqnarray}
\label{eq:FRW-tensor}
ds^2= a^2(\tau)\left[ -d\tau^2 + (\delta_{ij} + D_{ij}) dx^i dx^j \right] \,,
\end{eqnarray}
where $D_{ij}$ is a traceless divergence free tensor field.
Inserting Eq.~(\ref{eq:FRW-tensor}) into the field equation~(\ref{eq:fieldeq}), one can deduce the scale independent modified equation for the tensor mode evolution,
\begin{eqnarray}
\label{eq:tensor-evol}
D_{ij}^{\prime \prime} + \left( 2 \mathcal{H} + \frac{F_R^{\prime}}{F_R} \right) D_{ij}^{\prime} + k^2 D_{ij} =  \frac{a^2 \pi^T_{ij}}{M^2_{\text{Pl}} F_R}\,,
\end{eqnarray}
where $\mathcal{H} = a^{-1} da / d\tau$, the prime denotes the derivative with respect to the conformal time, and $\pi^T_{ij}$ is the tensor perturbation of $T_{ij}$.

\section{Constraints from cosmological observations}\label{sec:constraints}

In this section, we use the current observational data to constrain the $R^{2-q}$ inflation as well as the model parameters in $F(R)$ with dark energy.
We perform the CosmoMC package and the modified CAMB with the cosmological data including the CMB data from Planck
with both low-$l$ ($l < 50$) and high-$l$ ($l \ge 50$) parts
and WMAP with only the low-$l$ one as well as the BAO data from BOSS DR11 and the SNIa data from SNLS.

\begin{center}
\begin{figure}[ht]
\includegraphics[width=0.49 \linewidth]{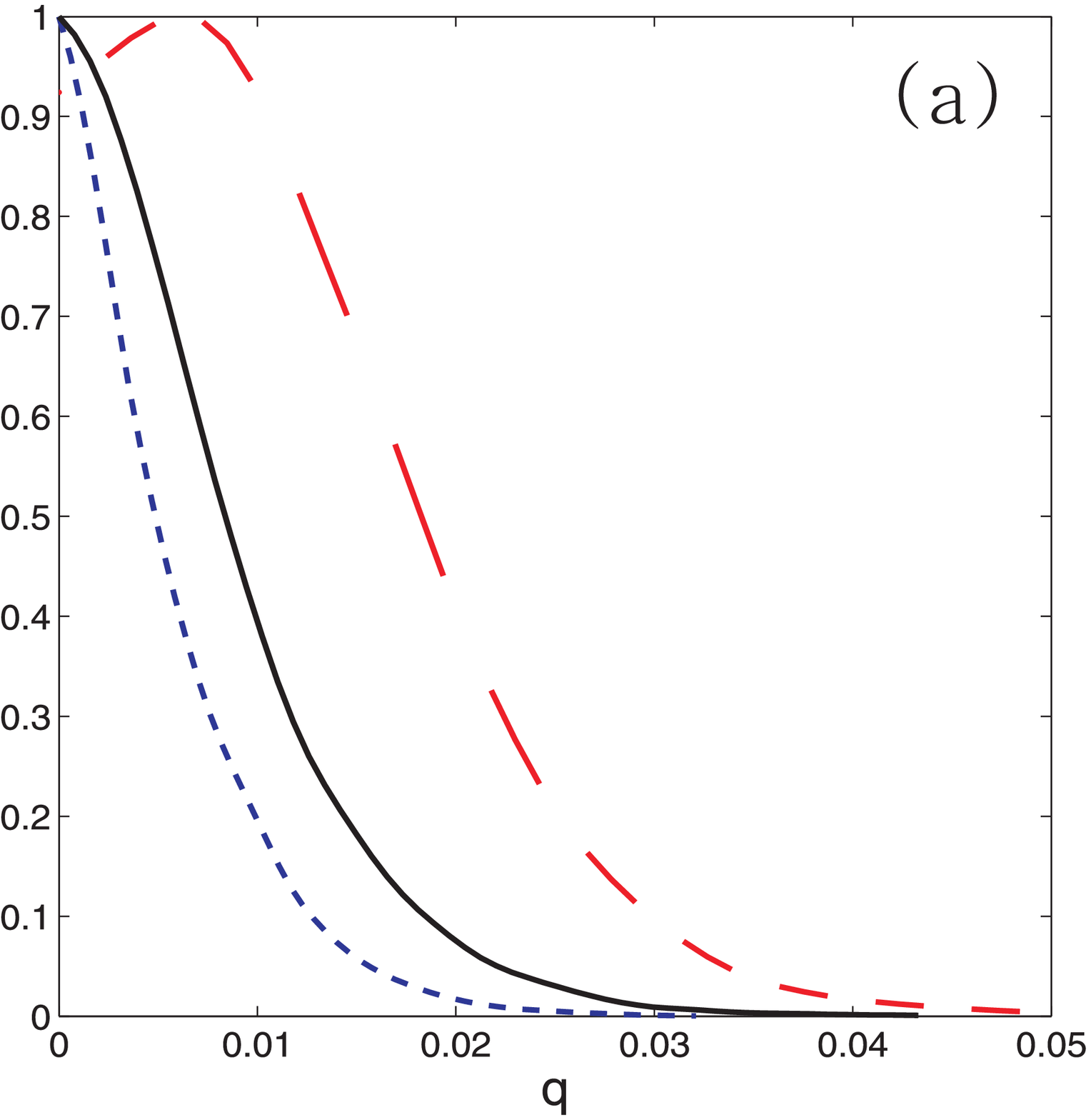}
\includegraphics[width=0.49 \linewidth]{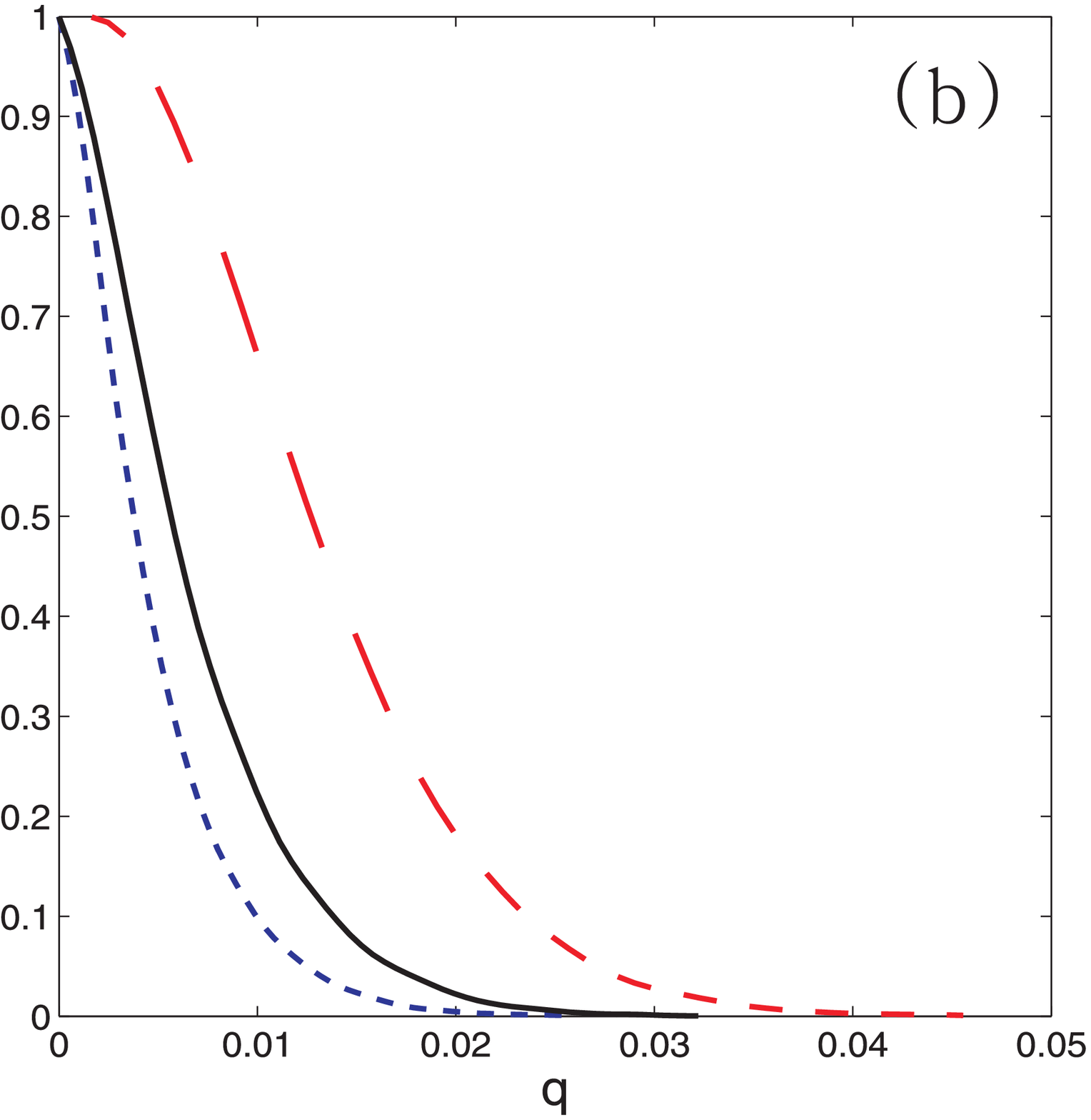}
\caption{
Marginalized probability for the inflation power parameter $q$ in (a) Starobinsky and (b) exponential gravity models, where the long-dashed, solid and dashed lines correspond to $N= 50,$ $60$ and $70$, respectively.
}
\label{fg:2}
\end{figure}
\end{center}

We modify two parts in the CAMB and CosmoMC packages. Firstly, we study the $R^{2-q}$ inflation model by adapting a standard running model to parametrize the primordial scalar and tensor perturbations,
\begin{eqnarray}
 \ln\mathcal{P}_s(k)& =& \ln A_s + (n_s -1) \ln \left(\frac{k}{k_s}\right) 
 \nonumber\\
 &&+ \alpha_s \left[\ln \left( \frac{k}{k_s}\right) \right]^2 \,, \\
\ln\mathcal{P}_t(k) &=&\ln A_t + n_t \ln \left( \frac{k}{k_s} \right) \,,
\end{eqnarray}
where $k_s$ is the pivot scale, $A_{s (t)}$ is the scalar (tensor) amplitude at the pivot scale with $A_t=r A_s$.
Secondly, we examine the scalar evolution by following the similar process of the MGCAMB code~\citep{Hojjati(2011)}, in which Eqs.~(\ref{eq:Poisson-1}) and (\ref{eq:Poisson-2}) are used to incorporate the $F(R)$ effect. Moreover, the tensor modification in Eq.~(\ref{eq:tensor-evol}) is also considered in our program.

In Fig.~\ref{fg:2}, we depict the $1D$ marginalized probability plot for the inflation power $q$ with 
(a) Starobinsky and (b) exponential gravity models.
As estimated in Sec.~\ref{subsec:rp_inf}, the $2-\sigma$ allowed region indicates $q \sim O(10^{-2})$. 
In particular, we have $q <2.66 \times 10^{-2}$ and $2.17 \times 10^{-2}$ in Starobinsky and exponential $F(R)$ models, respectively. 

\begin{center}
\begin{figure}[ht]
\includegraphics[width=0.49 \linewidth]{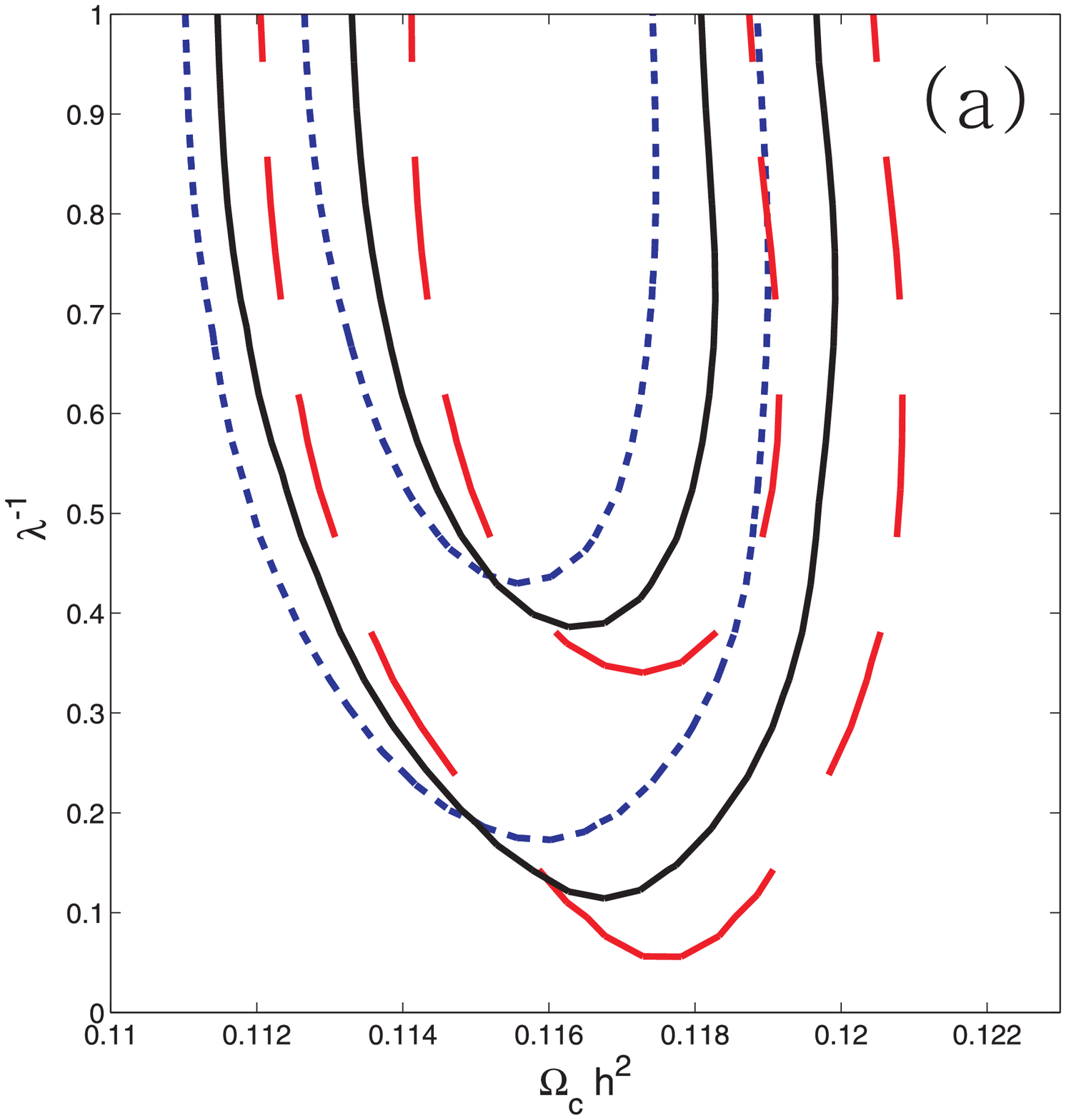}
\includegraphics[width=0.49 \linewidth]{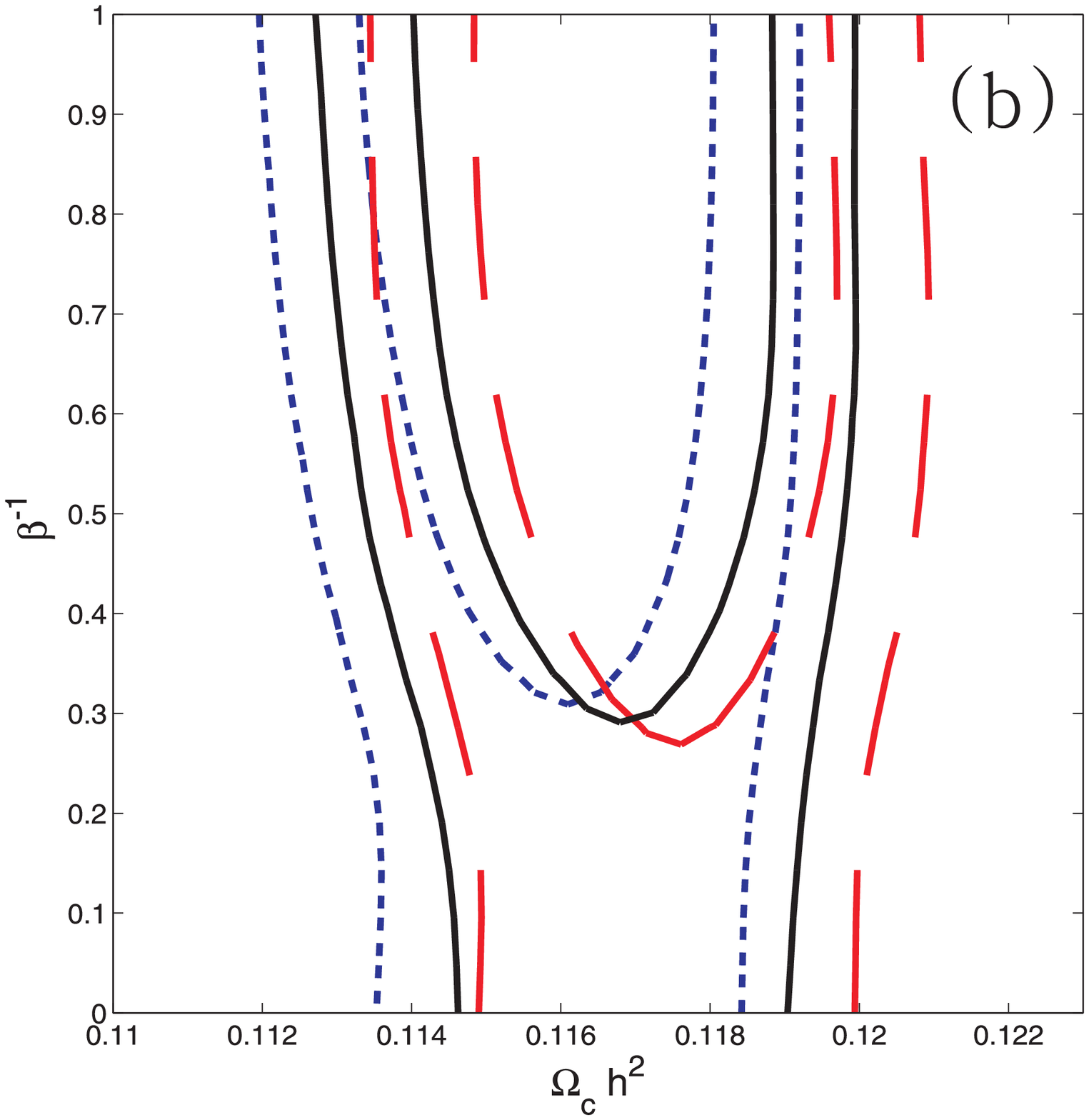}
\caption{
Contour plots in the planes (a) $\lambda^{-1}-\Omega_ch^2$ for Starobinsky and (b) $\beta^{-1}-\Omega_ch^2$ for exponential gravity models, where the inner and outer curves represent $1-\sigma$ and $2-\sigma$ confidence levels, while the long-dashed, solid and dashed lines correspond to $N= 50,$ $60$ and $70$, respectively.}
\label{fg:3}
\end{figure}
\end{center}

It is well-known that the $F(R)$ gravity exists a $\Lambda$CDM limit at $\lambda$ $(\beta) \rightarrow \infty$ with $\lambda R_c$  ($\beta R_c$) being equal to the cosmological constant $\Lambda$.
We examine all the allowed parameter space when we fit the model parameters with the prior $0<\lambda^{-1}$ $(\beta^{-1})<1$.
The constraints on the model parameter $\lambda^{-1}$ ($\beta^{-1}$) versus the physical density of cold dark matter (CDM) $\Omega_ch^2$ are shown in Fig.~\ref{fg:3}a (b), where 
the long-dashed, solid and dashed lines represent the e-foldings of $N= 50,$ $60$ and $70$, respectively.
We note that the best-fit of the Starobinsky (exponential) gravity model parameter locates 
around $\lambda^{-1} \simeq 0.75 $ ($\beta^{-1} \simeq 0.87$) with $\Omega_ch^2$ similar to the range in the $\Lambda$CDM model.
These results indicate that the evolution history in the Starobinsky model prefers a significant deviation from that in the standard cosmology, so that the $\Lambda$CDM limit with $\lambda^{-1} \rightarrow 0$ has been excluded within $2-\sigma$ confidence level.
On the contrary, the exponential type model behaves a smaller deviation from the standard cosmology~\citep{Geng(2014)}. 
\begin{table*}[ht]
\begin{center}
\caption{Allowed regions of the baryon density $\Omega_{b}h^2$, CDM density $\Omega_ch^2$, neutrino mass sum $\Sigma m_{\nu}$, 
inflation power parameter $q$, spectral index $n_s$, tensor-to-scalar ratio $r$, and $\sigma_8$
with $95\%~CL$  and $\xi\equiv 1-\lambda^{-1}\,(1-\beta^{-1})$ in the Starobinsky (exponential) model
 with $68\%~CL$, 
 where the difference of the best $\chi^2$ fit  between $f(R)$ and $\Lambda$CDM
 is defined by $\Delta \chi^2=\chi^2_{f(R)}-\chi^2_{\Lambda CDM}$.} 
\begin{tabular}{|c||c|c|c|c|c|c|c|} \hline
Parameters
&\multicolumn{3}{c|}{Starobinsky}&\multicolumn{3}{c|}{Exponential} & 
$\Lambda$CDM
\\ \cline{2-7}
{}&N=50&N=60&N=70&N=50&N=60&N=70&{}\\ 
 \hline
$100 \Omega_{b}h^2$& $2.24\pm 0.05$ & $2.25^{+0.04}_{-0.05}$ & $2.25\pm 0.05$ & $2.22 ^{+0.06}_{-0.04}$ & $2.24^{+0.04}_{-0.05}$ & $2.22^{+0.07}_{-0.03}$ & $2.20^{+0.06}_{-0.03}$ 
\\ \hline
$\Omega_ch^2$& $0.118^{+0.002}_{-0.05}$ & $0.118^{+0.001}_{-0.05}$ & $0.116^{+0.002}_{-0.04}$ & $0.119^{+0.001}_{-0.05}$ & $0.118^{+0.001}_{-0.004}$ & $0.117^{+0.001}_{-0.004}$ & $0.118 \pm 0.003$
\\ \hline
$\Sigma m_{\nu}$/eV& $<0.302$ & $0.063^{+0.266}_{-0.063}$ & $0.109^{+0.246}_{-0.109}$ & $<0.239$ & $<0.264$ & $0.086^{+0.211}_{-0.086}$ & $<0.211$
\\ \hline
$100 q$& $<2.66 $ & $<1.86$ & $<1.37$ & $<2.17$ & $<1.43$ & $<1.10$ & --
\\ \hline
$n_s$ & $0.972^{+0.005}_{-0.010}$ & $0.970^{+0.009}_{-0.002}$ & $0.973^{+0.008}_{-0.001}$ & $0.963^{+0.012}_{-0.001}$ & $0.970^{+0.007}_{-0.002}$ & $0.974^{+0.006}_{-0.002}$ & $0.963^{+0.012}_{-0.009}$
\\ \hline
$10^{3}r$ & $<11.1$ & $<6.69$ & $<4.46$ & $<9.42$ & $<5.63$ & $<3.93$ & $<125$
\\ \hline
$\sigma_8$ & $1.140^{+0.030}_{-0.139}$ & $1.131^{+0.038}_{-0.133}$ & $1.111^{+0.051}_{-0.117}$ & $0.962^{+0.021}_{-0.146}$ & $0.963^{+0.021}_{-0.151}$ & $0.940^{+0.042}_{-0.135}$ & $0.833^{+0.024}_{-0.059}$
\\ \hline
$\xi$
& $0.289^{+0.161}_{-0.289}$ & $0.287^{+0.111}_{-0.287}$ & $0.258^{+0.116}_{-0.258}$ & $0.127^{+0.348}_{-0.127}$ & $0.132^{+0.331}_{-0.132}$ & $0.197^{+0.263}_{-0.197}$ & --
\\ \hline
$\Delta \chi^2$ & $-2.65$ & $-2.61$ & $-1.74$ & $-1.32$ & $-0.88$ & $0.26$ &  --
\\ \hline
\end{tabular}
\label{table1}
\end{center}
\end{table*}

\begin{center}
\begin{figure}[ht]
\includegraphics[width=0.49 \linewidth]{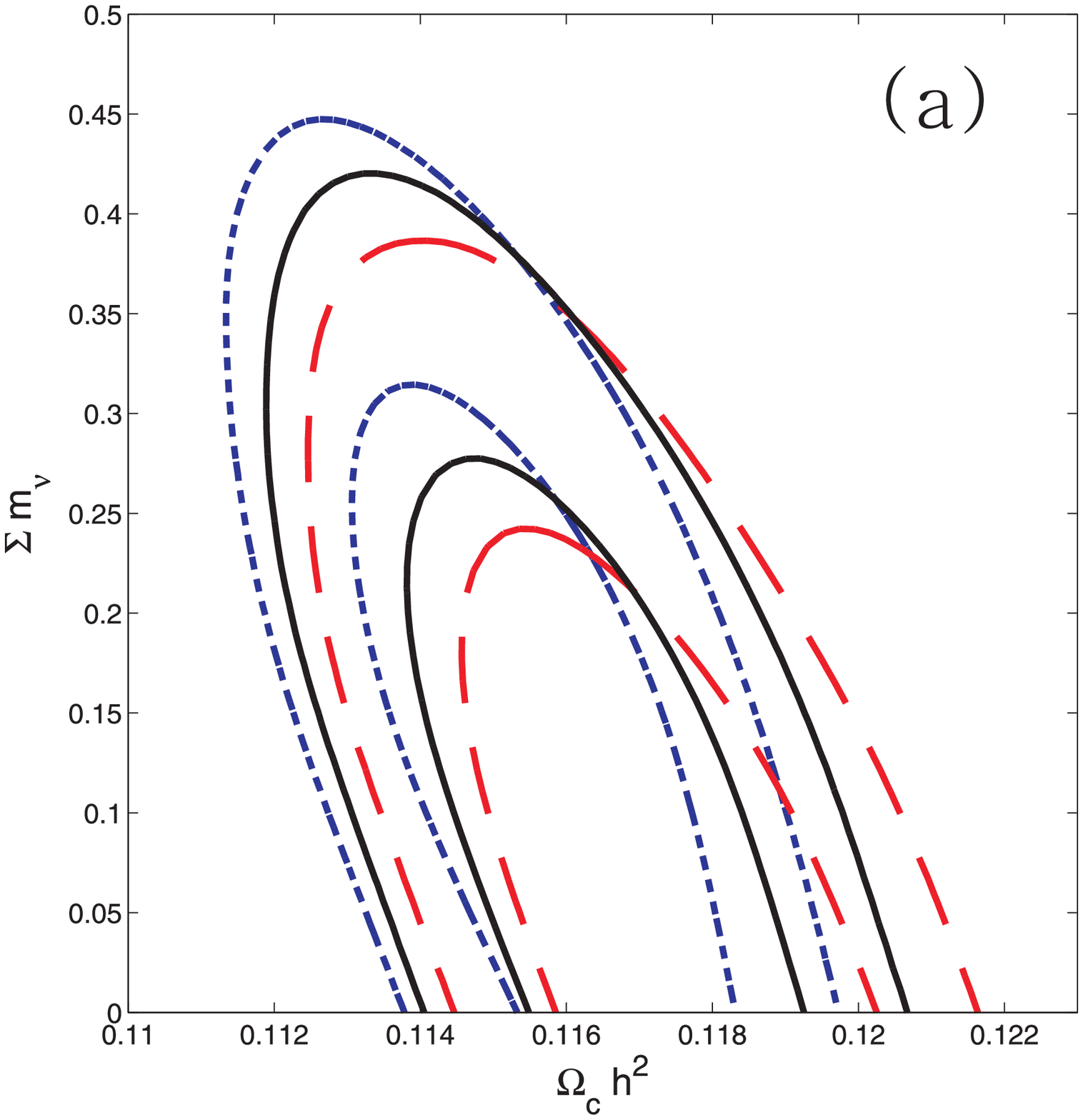}
\includegraphics[width=0.49 \linewidth]{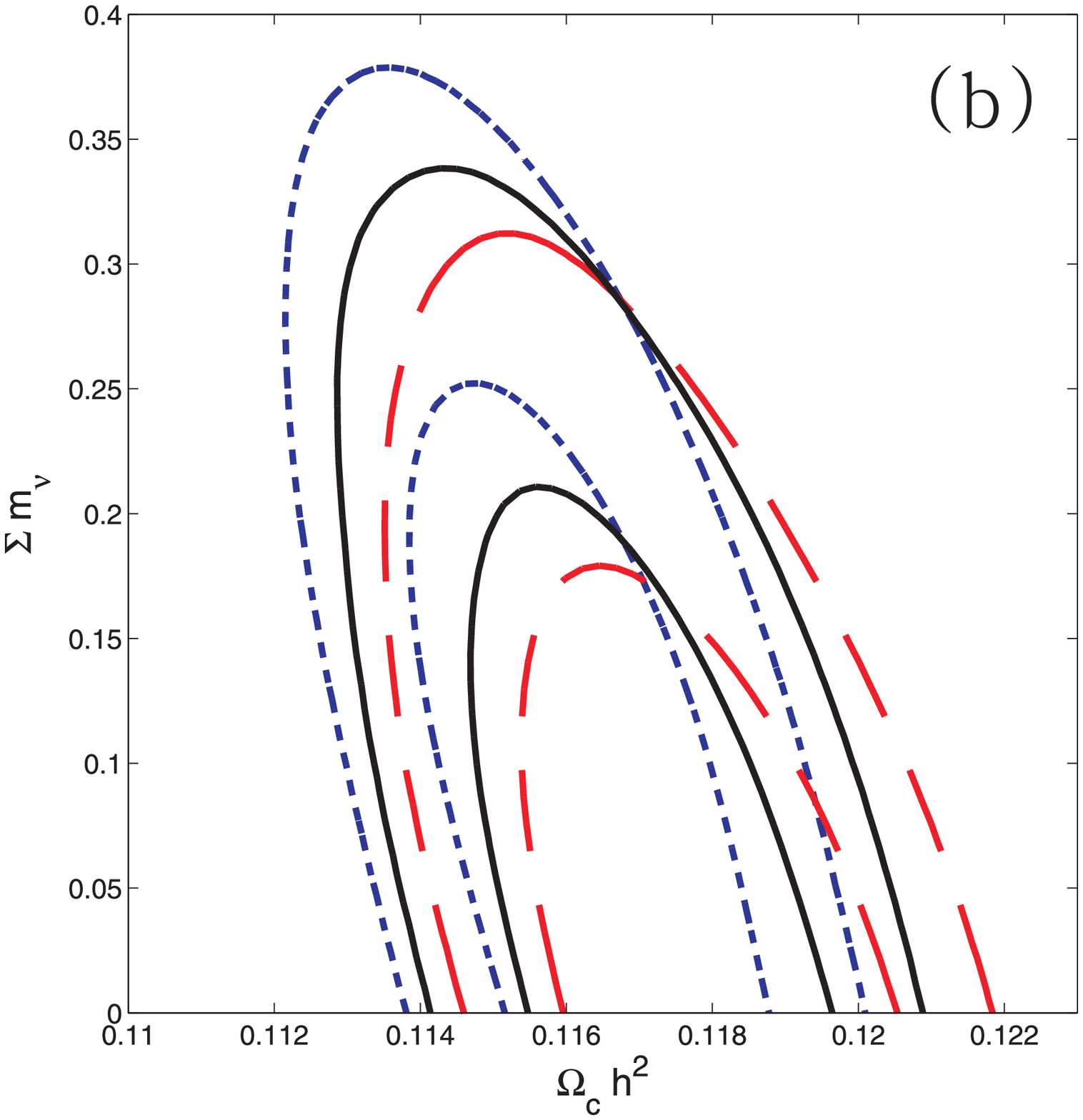}
\caption{
Contour plots in the planes $\Sigma m_{\nu}-\Omega_ch^2$ for (a) Starobinsky and (b) exponential gravity models, where the inner and outer curves represent $1-\sigma$ and $2-\sigma$ confidence levels, while the long-dashed, solid and dashed lines correspond to $N= 50,$ $60$ and $70$, respectively.}
\label{fg:4}
\end{figure}
\end{center}

Since the massive neutrinos play an important role in not only the cosmological evolutions but also particle physics experiments, we present the total neutrino mass $\Sigma m_{\nu}$ versus the dark matter density $\Omega_{c}h^{2}$ in Fig.~\ref{fg:4}.
In the Starobinsky (exponential) model, the allowed neutrino mass is given by $\Sigma m_{\nu} \lesssim 0.35~(0.30)$~eV. 
Comparing to the $\Lambda$CDM result of $\Sigma m_{\nu}<0.211$~eV, 
the $F(R)$ gravity results in a significant enhancement on the neutrino masses, especially the Starobinsky model.
Our results for the combined $f(R)$ scenarios are summarized in Table.~\ref{table1}.
Due to the $\chi^2$ difference, listed in Table.~\ref{table1}, we observe that the viable $f(R)$ gravity models 
in most of the examples perfectly fit the cosmic data ($\chi^2<0$), especially those with the smaller e-folding $N$.

\section{Conclusions} \label{sec:conclusion}

We have investigated the combined $f(R)$ models to unify inflation and dark energy scenarios. The inflation model is based on the generalized Starobinsky $f(R)$ inflation theory, $R^{2-q}$, while the late-time dark energy is described by the viable $f(R)$ gravity, including the Starobinsky and exponential viable $f(R)$ gravity, belonging to the power-law and exponential types of modified gravity models, respectively. 
In order to examine the consistency of the slow-roll parameters in the $R^{2-q}$ inflation and the combined $f(R)$ models, we numerically plot the potentials
 in both models and find that there is no difference between these two models in the inflationary stage but not in the post inflationary epoch.
The difference might cause a change in the reheating history.
In addition, the behavior of the curvature in the Jordan frame is also changed: the curvature in the $R^2$ inflation oscillates around $R=0$. 
However, when we include the viable $f(R)$ dark energy, the curvature keeps positive-definite and is forbidden to cross the critical curvature $R_{cr}$.

The constraints from the observational data 
is perfectly fitted by the combined model ($\chi^2<0$) and
point out that
although the $R^{2-q}$ inflation can loosen the tensor-to-scalar ratio bound in the $R^2$ inflation, $r$ is still restricted to a tiny value by the accurate observations, resulting in $q \lesssim 3 \times 10^{-2}$, which is hardly to be distinguished from the standard $R^2$ inflation.
We have also examined the model parameter of $\lambda$ ($\beta$) of the viable Starobinsky (exponential) $f(R)$ gravity to test the allowed parameter space. 
Although there exists a $\Lambda$CDM limit at $\lambda^{-1}$ ($\beta^{-1}$) $\rightarrow 0$, the Starobinsky model prefers $\lambda^{-1} \sim 0.7$, so that the $\Lambda$CDM limit is excluded within $2-\sigma$ confidence level, while the best-fit value in the exponential gravity locates at $\beta^{-1} \sim 0.85$, which deviates from the $\Lambda$CDM model within $1-\sigma$ confidence level.
We have also shown the deviations of the neutrino mass sum $\Sigma m_{\nu}$ between $f(R)$ gravity and the $\Lambda$CDM model.
Explicitly, we have found that the allowed $\Sigma m_{\nu}$ could be released to $0.36$ and $0.30$~eV 
in Starobinsky and exponential $F(R)$ dark energy models, respectively.

\begin{acknowledgments}
The work was supported in part by National Center for Theoretical Sciences, National Science
Council (NSC-101-2112-M-007-006-MY3) and National Tsing Hua
University (104N2724E1).
\end{acknowledgments}

\bibliographystyle{spr-mp-nameyear-cnd}


\end{document}